\documentclass[aps,pre,reprint,10pt,superscriptaddress,showpacs]{revtex4-1}
\usepackage{amsfonts} 
\usepackage{amsmath}
\usepackage{amssymb}
\usepackage{graphicx}
\usepackage{subfigure}
\usepackage{color}
\begin{document}
\title{Magnetohydrodynamic shocks in a  dissipative quantum plasma with exchange-correlation effects }
\author{ Biswajit Sahu}
\email{biswajit_sahu@yahoo.co.in}
\affiliation{Department of Mathematics, West Bengal State University,
Barasat, Kolkata-700126, India}
\author{A. P. Misra}
\email{apmisra@visva-bharati.ac.in; apmisra@gmail.com}
\affiliation{Department of Mathematics, Siksha Bhavana, Visva-Bharati University, Santiniketan-731 235, West Bengal, India}
\pacs{52.25.Xz, 52.35.Bj, 52.35.Tc}
 

%
%
\begin{abstract}
 We investigate the nonlinear propagation of multidimensional magnetosonic shock waves (MSWs) in a dissipative quantum magnetoplasma. A macroscopic quantum magnetohydrodynamic (QMHD) model is used to include the quantum force associated with the Bohm potential, the pressure-like spin force,  the exchange and correlation force of electrons, as well as  the dissipative force due to the kinematic viscosity of ions and the magnetic diffusivity.  The effects of these forces  on the properties of arbitrary amplitude MSWs are examined numerically.  It is found that the contribution from the exchange-correlation force appears to be dominant   over those from the pressure gradient and the other similar quantum forces, and it results into  a transition from  monotonic to oscillatory   shocks in presence of either the ion kinematic viscosity or the  magnetic diffusivity. 
\end{abstract}
\maketitle 
 
\section{Introduction}\label{intro}
Quantum plasmas have received a considerable attention over the last decade  as a means of its potential applications in   solid state physics, in microelectronics \cite{mic}, in superdense
astrophysical systems (particularly, in the interior of Jupiter, white dwarfs and superdense neutron stars) \cite{ast,spac1,spac2} in nano-particles, quantum-wells, quantum-wires, and quantum-dots
\cite{nan}, in ultracold plasmas \cite{ult},   in carbon nanotubes and quantum diodes \cite{wel}, in nonlinear optics \cite{opt},    in high-intensity laser-produced plasmas \cite{las1}-\cite{las3} etc.  
\par
Recently, there has been a growing and considerable interest in investigating new aspects of quantum plasma physics by developing non-relativistic quantum hydrodynamic (QHD) model
\cite{qhd1}-\cite{qhd3}. The QHD model generalizes the fluid model of plasmas with inclusion of a quantum correction term known as Bohm potential in momentum transfer equation to describe quantum
diffraction effects. Moreover, quantum statistical effects appear in the QHD model through an equation of state. The collective motion of quantum particles in magnetic fields gives rise an
extension to the classical theory of magnetohydrodynamics (MHD) in 
terms of the well-known quantum magnetohydrodynamics (QMHD) \cite{qhd4}. The QMHD plasmas are  of importance in astrophysical plasmas, such as neutron stars, pulsar magnetosphere, magnetars etc. From the laboratory perspective, the motion of particles with spin effects are important under strong magnetic fields as a probe of quantum physical phenomena \cite{mhd1}-\cite{mhd3}. Furthermore, for quantum systems the interactions between electrons can be separated into a Hartree term due to the electrostatic potential of the total electron density and an electron exchange-correlation term because of the electron-$1/2$ spin effect. When the electron density is high, and the electron temperature is low,   the
electron exchange-correlation effects, in particular, should be important \cite{exc1}. These forces in the collective behaviors of   plasmas play    crucial roles on the nonlinear wave dynamics \cite{exc2}-\cite{exc6}.
\par
 Furthermore, the concept of spin MHD is important when the difference in energy between two spin states is larger than the thermal energy and the presence of large number of particles in the Debye sphere does not necessarily influence the importance of spin effects \cite{mar1}. Marlkund and Brodin \cite{bro1} have recently extended the QMHD model to include the spin-magnetization effects by introducing a generalized term for the so-called quantum force. It was found that the collective spin effects may influence the propagation characteristics of nonlinear waves in a strongly magnetized quantum plasma. It has been shown that  the typical plasma behaviors can be significantly changed by the electron spin properties and the plasma can even show
ferromagnetic behaviors in the low-temperature and high-density regimes \cite{bro2}. 
\par
On the other hand, nonlinear magnetosonic waves (MWs) in the classical regime have been investigated due to their importance in space, astrophysical  and fusion plasmas, with application to particle heating and acceleration. Nonlinear collective processes in quantum plasmas have also been studied by including both the quantum tunneling and the electron spin effects on an equal footing, which can give rise to new collective linear and nonlinear magnetosonic excitations.  Marklund \textit{et al.} \cite{mar2} studied magnetosonic solitons in a non-degenerate quantum plasma with the Bohm potential and electron spin-$1/2$ effects. Misra and Ghosh \cite{mis1} investigated the small amplitude MWs in a quantum plasma taking into account the effects of the quantum tunneling and the electron spin. Recently, Mushtaq and Vladimirov \cite{mus1} studied the magnetosonic solitary waves in spin-$1/2$ quantum plasma. They incorporated the  spin effects by taking into account the spin force and the macroscopic spin magnetization current.
 However, most of these investigations are limited to one-dimensional (1D) planar geometry which may not be a realistic situation in laboratory devices, since the waves observed in laboratory devices are certainly not bounded in one-dimension, and do not consider the effects of the exchange-correlation force as well as the plasma resistivity and the viscosity effects together. 
 \par
 The purpose of the present work is to consider these quantum and the dissipative effects consistently, and to study the nonlinear propagation of multidimensional arbitrary amplitude  magnetosonic shock waves (MSWs) in spin quantum magneto-plasmas. We show that the exchange correlation force, which was omitted in the previous studies \cite{sahu2015}, plays a dominating role over other similar forces  on the formation of monotonic and  oscillatory  MSWs.
\section{Theoretical model }   We consider the nonlinear propagation of large amplitude QMHD waves in a dissipative magneto-plasma  consisting of quantum electrons and classical viscous ions. The QMHD equations  
 for electrons   are \cite{bro1,misra2010}
\begin{equation}
\frac{\partial n_{e}}{\partial t}+\nabla \cdot\left(
n_{e}\mathbf{v}_{e}\right) =0  \label{e1}
\end{equation}
\begin{equation}
\frac{d \mathbf{v}_{e}}{dt} =-\frac{e}{%
m_{e}}\left( \mathbf{E}+\mathbf{v}_{e}\times \mathbf{B}\right)
-\frac{\nabla
P_{e}}{m_{e}n_{e}}+\frac{\mathbf{C}_{ei}}{m_{e}n_{e}}+\mathbf{F}_{q},
\label{e2}
\end{equation}
\begin{equation}
\frac{d \mathbf{S}}{dt} =-\frac{2\mu }{%
\hbar }\left( \mathbf{B}\times \mathbf{S}\right) ,  \label{e3}
\end{equation}
where $d/dt\equiv\partial_t+\textbf{v}_e\cdot\nabla$, $\mathbf{C}_{ei}$ represents the collisions between electrons
and ions and $\mathbf{F}_{q}$ is the total quantum force given by
\begin{equation}
\mathbf{F}_{q}=\frac{\hbar ^{2}}{2m_{e}^{2}}\nabla \left( \frac{\nabla ^{2}%
\sqrt{n_{e}}}{\sqrt{n_{e}}}\right) +\frac{2\mu }{m_{e}\hbar }\mathbf{S}%
\cdot\nabla \mathbf{B}+\frac{1 }{m_{e}} \nabla V_{xc}, \label{e4}
\end{equation}
in which the first term is associated with the Bohm potential (particle dispersion), the second term is the pressure-like spin force and the third one is associated with the exchange-correlation potential $V_{xc}$, given by \cite{exc1,exc3,hedin1971}
\begin{equation}
V_{xc}\approx
0.985(3\pi^2)^{2/3}\left(\frac{\hbar^2\omega_{pe}^2}{m_eV_{Fe}^2}\right)\left(\frac{n_e}{n_{0}}\right)^{1/3}.  
\label{eq5}
\end{equation}
In Eqs. \eqref{e1}-\eqref{eq5}, $m_j$, $n_j$,   $\mathbf{v}_j$ and $P_j$, respectively, denote the mass, number density, velocity and thermal pressure of $j$-species particles, where $j=e~(i)$ stands for electrons (ions). Also,  $\textbf{E}~(\textbf{B})$ is the electric (magnetic) field, $\textbf{S}$ is the spin angular momentum with $|\textbf{S}|=\hbar/2$ and  $\mu=-(g/2)\mu_B$ with $\hbar$ denoting the reduced Planck's constant, $g$   the electron $g$-factor and $\mu_B=e\hbar/2m_e$   the Bohr magneton. Furthermore,  $V_{Fe}\equiv
\sqrt{2k_BT_{Fe}/m_e}=(\hbar/m_e)(3\pi^2n_{0})^{1/3}$ is the Fermi velocity, where $k_B$ is the Boltzmann constant, $T_{Fe}$ is the electron Fermi temperature and $n_0$ is the equilibrium density of electrons and ions. The electromagnetic fields are coupled through the Maxwell's equations
\begin{equation}
\nabla \times \mathbf{E}=-\partial _{t}\mathbf{B},\hskip5pt \nabla\cdot%
\mathbf{B}=0,  \label{e5}
\end{equation}%
\begin{equation}
\nabla \times \mathbf{B}=\mu _{0}\left( \mathbf{j}_{D}+\mathbf{j}+\mathbf{j}%
_{M}\right),  \label{e6}
\end{equation}%
where $J$'s are the displacement current
$\mathbf{j}_{D}=\varepsilon
_{0}\partial _{t}\mathbf{E,}$ spin-magnetization current $\mathbf{j}%
_{M}=\nabla \times \mathbf{M=}\left( 2\mu /\hbar \right) \nabla \times n_{e}%
\mathbf{S}$ and the classical free current $\mathbf{j=}e\left( n_{i}\mathbf{v%
}_{i}\mathbf{-}n_{e}\mathbf{v}_{e}\right) $.
\par
The ion fluid equations read
\begin{equation}
\frac{\partial n_{i}}{\partial t}+\nabla .\left(
n_{i}\mathbf{v}_{i}\right) =0,  \label{e7}
\end{equation}%
\begin{eqnarray}
\left( \partial _{t}+\mathbf{v}_{i}.\nabla \right) \mathbf{v}_{i}=&&\frac{e}{
m_{i}}\left( \mathbf{E}+\mathbf{v}_{i}\times \mathbf{B}\right)-\frac{\nabla P_{i}}{m_{i}n_{i}} \notag\\
&&+\frac{\mathbf{C}_{ie}}{m_{i}n_{i}}+\frac{\zeta' }{
m_{i}n_{i}}\nabla ^{2}\mathbf{v}_{i},  \label{e8}
\end{eqnarray}
where  $\zeta'$ is the coefficient of the ion kinematic viscosity and $\mathbf{C}_{ie}$ is the collisions between ions and electrons. 
Defining the total mass density by $\rho =m_{e}n_{e}+m_{i}n_{i}$,  the
center-of-mass fluid velocity by $\mathbf{v}=\left( m_{e}n_{e}\mathbf{v}%
_{e}+m_{i}n_{i}\mathbf{v}_{i}\right) /\rho $,  the  set of reduced QMHD equations can be obtained from Eqs. \eqref{e1}-\eqref{e8} as 
\cite{bro1}
\begin{equation}
\frac{\partial \rho }{\partial t}+\nabla\cdot\left( \rho
\mathbf{v}\right) =0. \label{e21}
\end{equation}%
\begin{equation}
\left( \partial _{t}+\mathbf{v}\cdot\nabla \right) \mathbf{v}=\frac{1}{\mu _{0}\rho}%
(\nabla \times \mathbf{B}) \times \mathbf{B}-\frac{\nabla P}{\rho
}+\mathbf{F}_{q}+\zeta \nabla ^{2}\mathbf{v}, \label{e22}
\end{equation}%
\begin{equation}
\frac{\partial \mathbf{B}}{\partial t}=\nabla \times \left(
\mathbf{v}\times \mathbf{B}\right) +\lambda \nabla ^{2}\mathbf{B},
\label{e23}
\end{equation}%
where  $P\equiv P_{e}+P_{i}$ is the scalar pressure in the
center-of-mass frame, $\zeta=\zeta'/\rho $ is the coefficient of
ion kinematic viscosity,  $\lambda =\eta /\mu _{0}$ is the magnetic diffusivity and
\begin{equation}
\mathbf{F}_{q}=\frac{\hbar ^{2}}{2m_{e}m_{i}}\nabla \left( \frac{\nabla ^{2}%
\sqrt{\rho }}{\sqrt{\rho }}\right) +\frac{2\mu }{\hbar
m_{i}}\mathbf{S} +\frac{1 }{m_{i}}\nabla V_{xc}.  \label{e11}
\end{equation} 
In Eqs. \eqref{e21}-\eqref{e23}, we have used the MHD approximation, i.e., the quasineutrality condition, i.e., $n_{e}\approx n_{i},$ which gives $n_{e}=\rho /\left( m_{e}+m_{i}\right) \approx \rho /m_{i}$, $\mathbf{C}_{ei}=en_{e}\eta \mathbf{j}$,  where $\eta $ is the plasma resistivity,  and neglected the displacement current. Furthermore, we have considered the fact that in MHD, the scale lengths are typically $\gtrsim r_{L}$, the Larmor radius for ions. So, the terms
that are quadratic in $S$ can be neglected in the expression for the quantum force as well as in the spin-evolution equation. Also, to the lowest order, the spin inertia can be neglected for
frequencies well below the electron cyclotron frequency. Thus, we have for the spin-evolution equation $\mathbf{B}\times \mathbf{S=0}$, which gives
\begin{equation}
\mathbf{S=-}\frac{\hbar }{2}\tanh \left( \frac{\mu
_{B}B}{k_{B}T_{e}}\right) \mathbf{\hat{B}}.
\end{equation}
This expression of $\textbf{S}$ is to be substituted in $\textbf{F}_q$ [Eq. \eqref{e11}].
\par
In the appropriate dimensionless variables, Eqs. \eqref{e21}-\eqref{e23} 
can be recast in two space dimensions as
\begin{equation}
\frac{\partial \rho}{\partial t}+\frac{\partial }{\partial x}(\rho u)+\frac{%
\partial }{\partial y}(\rho v)=0,  \label{e24}
\end{equation}
\begin{eqnarray}
&&\left( \frac{\partial }{\partial t}+u\frac{\partial }{\partial x}+v\frac{%
\partial }{\partial y}\right) (u,v)=-\frac{B}{\rho} \frac{\partial B}{\partial (x,y)}\notag\\
&&-c_s^2 \frac{\partial }{\partial (x,y)}(ln \rho)
+\beta \frac{\partial }{\partial (x,y)}\left[\frac{1}{\sqrt{\rho}}\left(\frac{\partial
^{2}\sqrt{\rho}}{\partial x^{2}}+\frac{\partial
^{2}\sqrt{\rho}}{\partial y^{2}}\right)\right]\notag \\
&&+\frac{\varepsilon}{v_B^2 \rho} \frac{\partial }{\partial
(x,y)}\left[\rho B \tanh(\varepsilon B)\right] 
+\alpha \frac{\partial }{\partial (x,y)} \rho^{1/3} \notag \\
&&+\delta \left( \frac{\partial^{2}}{\partial x^{2}}+\frac{\partial ^{2}}{\partial y^{2}}\right)
(u,v),  \label{e25}
\end{eqnarray}
\begin{equation}
\frac{\partial B}{\partial t}+ \frac{\partial }{\partial x}(uB)+
\frac{\partial }{\partial y}(vB)-\gamma \left( \frac{\partial
^{2}}{\partial x^{2}}+\frac{\partial ^{2}}{\partial y^{2}}\right)
B=0, \label{e27}
\end{equation}%
where $\alpha= 0.985(3\pi^2)^{2/3}(m_e/m_i)H^2V_{Fe}^2/C_A^2$ with
$H=\hbar\omega_{pe}/m_e V_{Fe}^2$ denoting the ratio of electron plasmon energy to the Fermi energy densities, $\mathbf{B}$ is the magnetic
field along the $z-$axis, i.e., $\mathbf{B}=B(x,y,t)\hat{z}$,
normalized to its equilibrium value $B_0$. Also, the total mass density $\rho$ is  normalized to its equilibrium value $\rho_0$, the velocity $\mathbf{v}$ is  normalized to  the
Alfv{\'e}n speed $C_A=\sqrt{B_0^2/\mu_0 \rho_0}$. The space
and time variables are normalized to, respectively,
$C_A/\omega_{ci}$ and the ion gyroperiod $\omega_{ci}^{-1}$, where
$\omega_{ci}=eB_0/m_i$. Furthermore, $\beta=2 c^2(m_e/m_i)
\omega_{ci}^2\lambda_C^2/C_A^4$, where
$\lambda_C=c/\omega_C=\hbar/2m_e c$ is the Compton wavelength,
$\omega_C$ is the Compton frequency, $c$ is the speed of light in
vacuum, $c_s=\sqrt{k_B(T_e+T_i)/m_i}/C_A$ is the ion-acoustic  speed
normalized to $C_A$, $T_e(T_i)$ is the electron (ion) temperature,
and $k_B$ is the Boltzmann constant. Moreover, $v_B^2=k_BT_e/m_i
C_A^2=(1/\varepsilon)mu_B B_0/m_i C_A^2$ with
  $\varepsilon=\mu_BB_0/k_BT_e$ denoting the Zeeman energy, $\delta=\zeta \omega_{ci}/C_A^2$ is a dimensionless viscosity parameter and   $\gamma=\lambda \omega_{ci}/C_A^2$ is a dimensionless magnetic diffusivity parameter.
\section{Arbitrary amplitude Shocks }
We   consider the propagation of arbitrary amplitude stationary shock waves in
a planar geometry. In the moving frame of reference $\xi=l_x x+l_y
y-M t$, where $M$ is the Mach number and $l_x$ and $l_y$ are the direction cosines along the axes ($l_x^2+l_y^2=1$),  Eqs. \eqref{e24}-\eqref{e27} reduce to  a single differential equation in the magnetic field $B$ as
\begin{eqnarray}
     \frac{1}{2}\frac{d }{d \xi}\left(\frac{M }{\rho}\right)^2&&+\frac{B}{\rho} \frac{d B}{d \xi}
+c_s^2 \frac{d }{d \xi}(ln \rho)-\beta \frac{d }{d
\xi}\left(\frac{1}{\sqrt{\rho}}\frac{d^{2}\sqrt{\rho}}{d
\xi^{2}}\right) \notag\\
&&-\frac{\varepsilon}{v_B^2 \rho} \frac{d }{d \xi}\left[\rho B
\tanh(\varepsilon B)\right]-\alpha \frac{d }{d \xi}
\rho^{1/3}\notag\\
&&+\delta \frac{d^{2}}{d \xi^{2}}\left(\frac{M
}{\rho}\right)=0, \label{e32}
\end{eqnarray}
where
\begin{equation}
\rho=\frac{MB}{M-\gamma {dB}/{d\xi}}, \label{e33}
\end{equation}
and we have imposed the boundary conditions $\rho\rightarrow 1$,
$B \rightarrow 1$, $(u, v) \rightarrow (0,0)$, ${d \rho}/{d\xi}
\rightarrow 0$, ${d B}/{d\xi} \rightarrow 0$ as $|\xi|\rightarrow
\infty$.
\par 
 Equations \eqref{e32} and \eqref{e33} govern the evolution of arbitrary amplitude MSWs in a quantum plasma.  In Eq. \eqref{e32}, the contributions of different forces can be identified. The term $\propto c_s^2$ appears due to the thermal pressures of electrons and ions, the term $\propto \beta$ is due to the quantum particle dispersion associated with the Bohm potential, the contribution from the pressure-like spin force is $\propto \varepsilon$ and the term $\propto\alpha$ is the contribution from the exchange-correlation force of electrons. Furthermore, the terms $\propto\gamma$ and $\delta$ are from the dissipative effects due to the magnetic diffusivity and the ion kinematic viscosity respectively.
\section{Results and discussion}
\begin{figure*}
\includegraphics[width=0.6\textwidth]{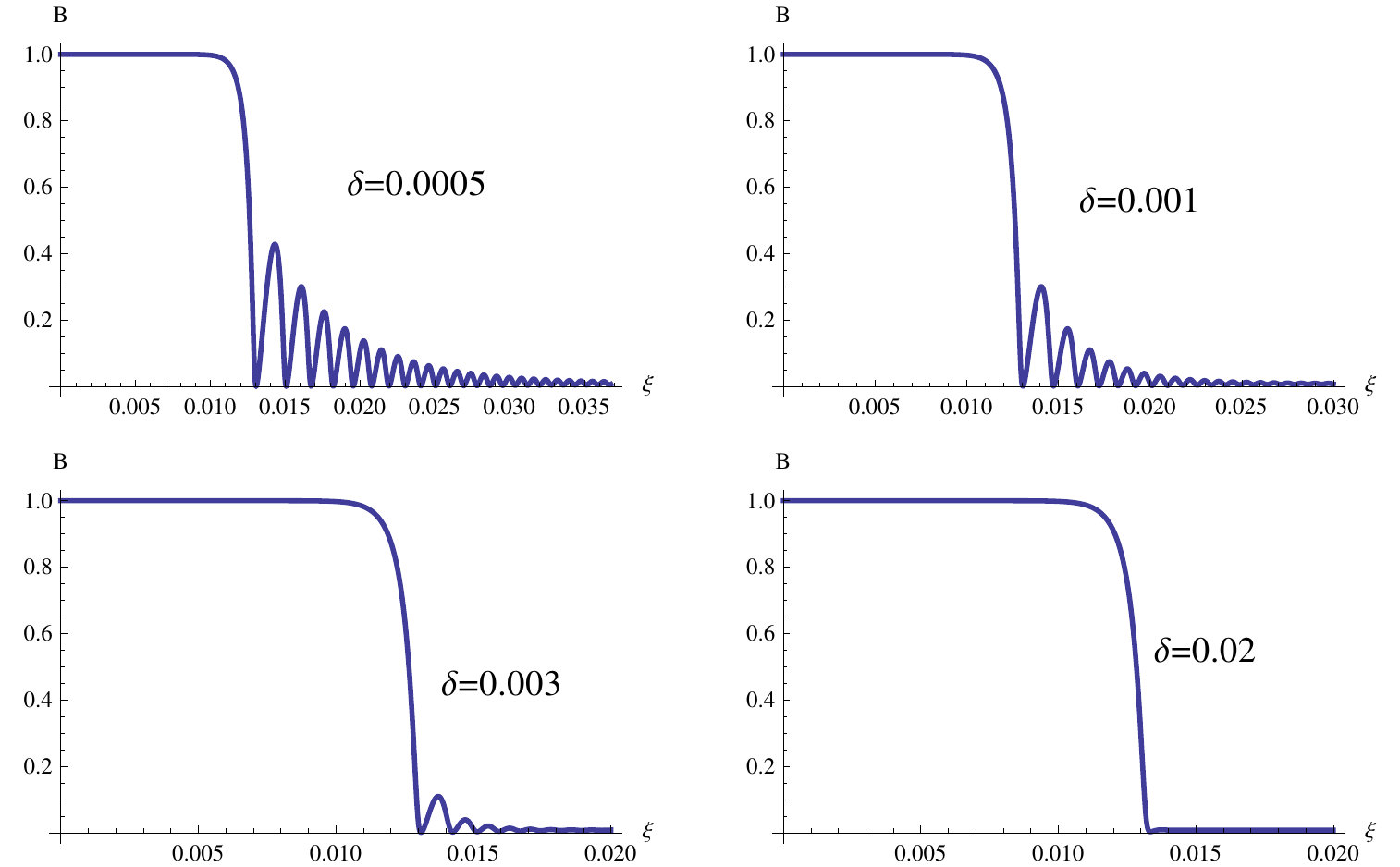}
\caption{Profiles of MSWs [solution of  Eq. \eqref{e32}] are shown for
different values of the viscosity parameter $\delta$ as in the figure. The other parameter values are  $n_0=4.0\times 10^{32} m^{-3}$, $B_0=5.0\times 10^{3}$ T, $T_e=5.0\times 10^{3}$ K, $T_i=0.1\times T_e$, $\gamma=0.001$ and $M=1.5$.}
\label{fig-1}
\end{figure*}

\begin{figure*}
\includegraphics[width=0.6\textwidth]{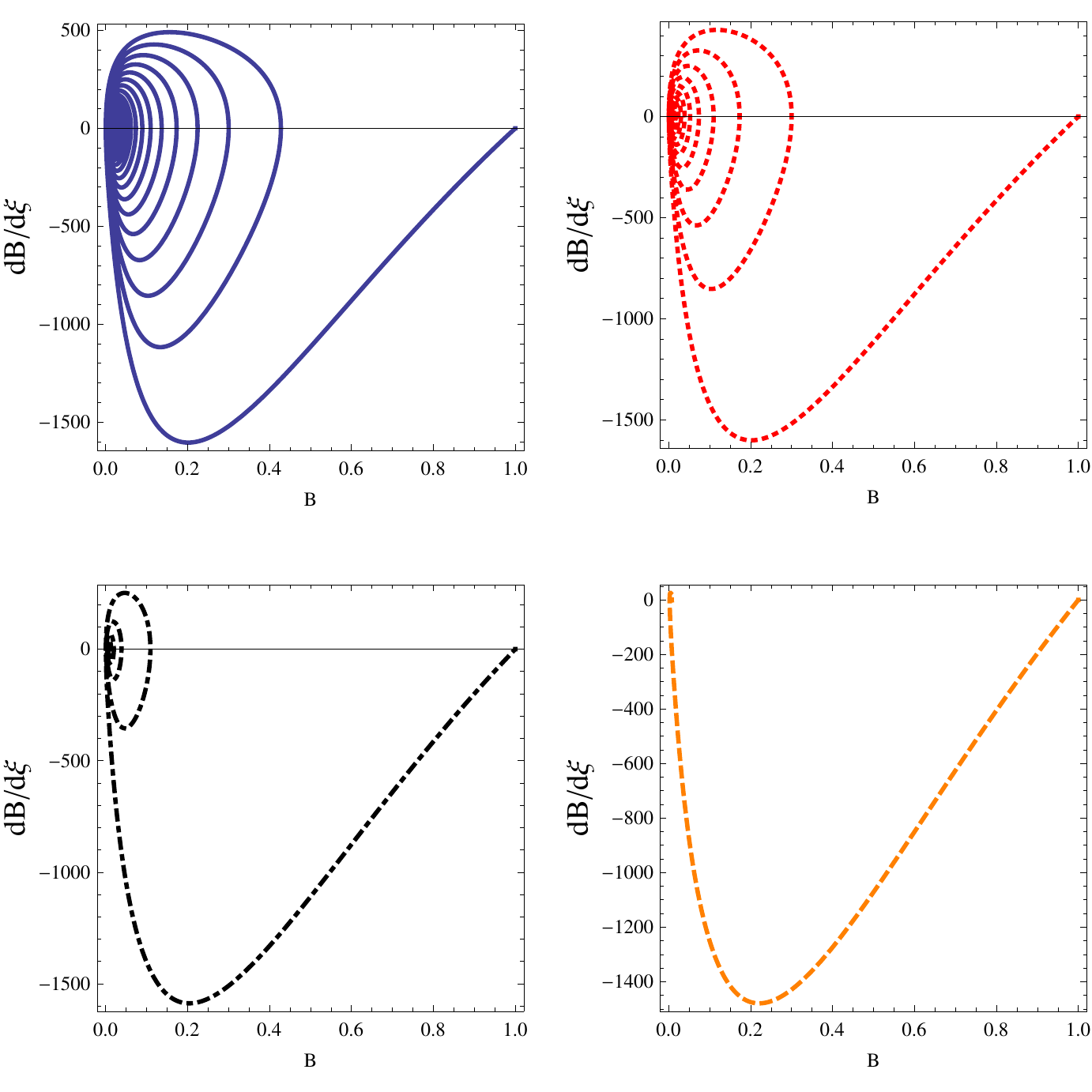}
\caption{The phase portraits [solution of  Eq. \eqref{e32}] are shown for different
values of $\delta$ as in Fig. \ref{fig-1}. The  solid, dotted, dash-dotted and dashed lines [from left to right (upper to lower) panels], respectively, correspond to $\delta=0.0005,~0.003,~0.001$ and  $0.02$. The other parameters are the same as in Fig. \ref{fig-1}.}
\label{figure2}
\end{figure*}

\begin{figure*}
\includegraphics[width=0.6\textwidth]{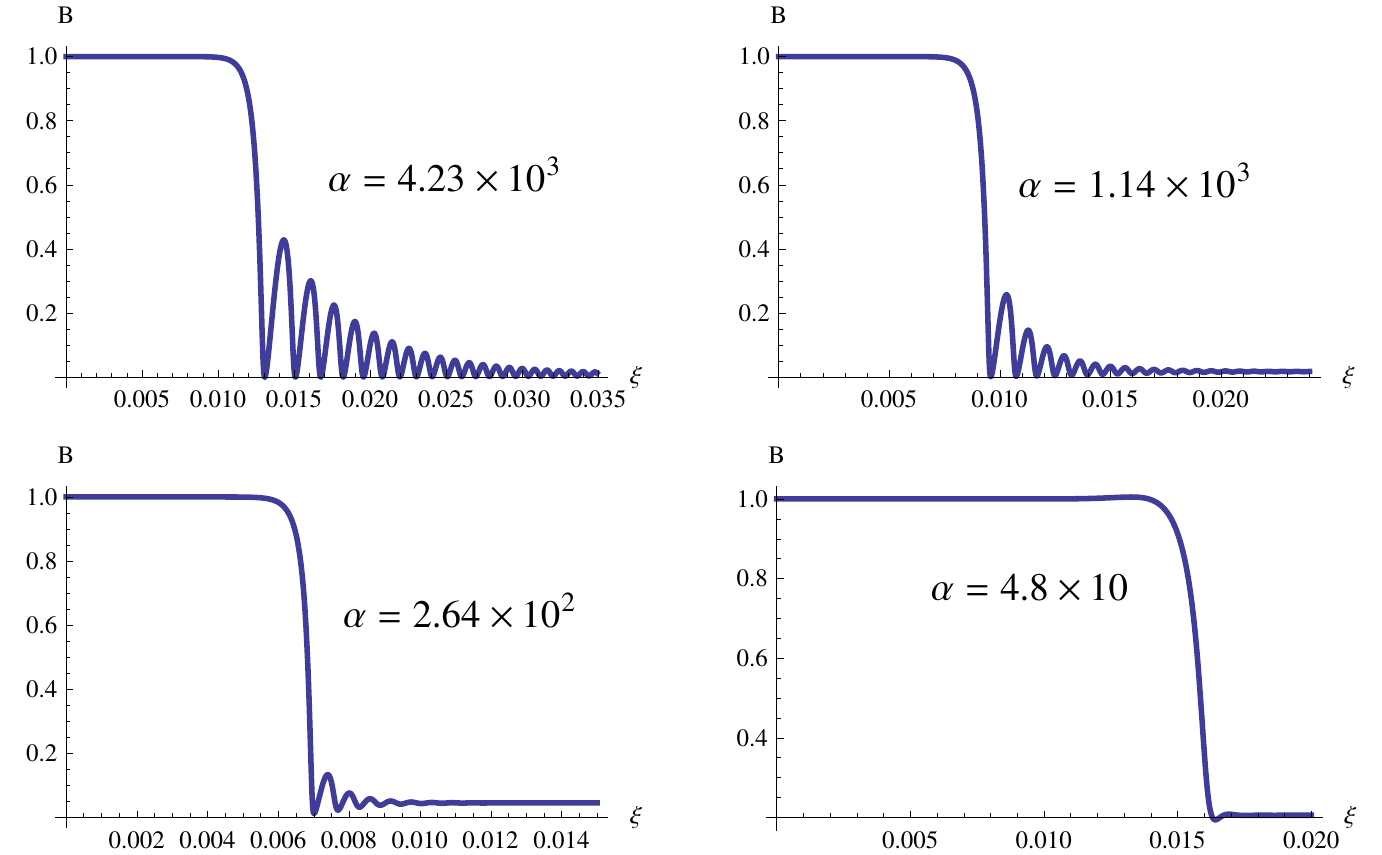}
\caption{Profiles of nonlinear solutions of Eq. (\ref{e32}) for
different values of exchange correlation parameter $\alpha$
($\alpha=4.23\times 10^{3}$ and $H=0.19$ for $n_0=4.0\times
10^{32}$, $\alpha=1.14\times 10^{3}$ and $H=0.22$ for
$n_0=1.5\times 10^{32}$, $\alpha=2.64\times 10^{2}$ and $H=0.27$
for $n_0=5.0\times 10^{31}$, $\alpha=4.8\times 10^{1}$ and
$H=0.33$ for $n_0=1.4\times 10^{31}$), where $\delta=0.0005$, and
the other parameters are same as in Fig. \ref{fig-1}.}
\label{fig-3}
\end{figure*}

In this section, we numerically investigate the properties of magnetosonic shocks which are    solutions of  Eq. \eqref{e32}. The profiles of the magnetic field are exhibited graphically in Figs. \ref{fig-1}-\ref{fig-3} for different values of the plasma parameters.  We note that the nature of shocks depends on the competition between the nonlinearity (causing wave steepening) and the dissipation (causing wave energy to decay) of the medium. When the wave breaking due to
nonlinearity is balanced by the combined effects of dispersion and dissipation, a monotonic or oscillatory   shocks are generated in a plasma \cite{shuk}. On the other hand, if the dissipation in the system is small, the particle trapped in a potential well will fall to the bottom of the well  while performing oscillations between its wall, and one obtains an oscillatory wave. For very
small values of the dissipation in the system, the energy of the particle decreases slowly, and the first few oscillations at the wave front will be close to solitons. Furthermore, if the contribution from the dissipation is larger than its critical value,   the motion of the particle will be aperiodic and monotonic shock structures will be formed. 
\par
Inspecting the magnitudes of the coefficients of Eq. \eqref{e32} we find that for non-relativistic quantum plasmas, 
\begin{equation}
\frac{\beta}{\alpha}\sim \frac{m_e}{m_i}\frac{V_{Fe}^2}{c^2}\ll1;~\frac{\varepsilon^2/v_B^2}{c_s^2}\sim\left(\frac{\mu_BB_0}{k_BT_e}\right)^2\left(\frac{m_iC_A^2}{k_BT_e}\right)^2, \label{estimation} 
\end{equation}
\begin{eqnarray}
\frac{\varepsilon^2/v_B^2}{\alpha}&&\sim0.1\left(\frac{\mu_BB_0}{k_BT_e}\right)^2 \frac{m_iC_A^2}{k_BT_e} \frac{m_i}{m_e}\frac{C_A^2}{H^2V_{Fe}^2}\notag\\
&&>0.1 \left(\frac{\mu_BB_0}{k_BT_e}\right)^2\left(\frac{m_iC_A^2}{k_BT_e}\right)^2 \notag\\
&&\sim 0.1\frac{\varepsilon^2/v_B^2}{c_s^2}, i.e., \alpha\gg c_s^2. \label{estimation2} 
\end{eqnarray}
Thus, from Eqs. \eqref{estimation1} and \eqref{estimation2} it follows that   the contributions from the pressure gradient and the spin forces may be comparable, however, the contribution from the exchange-correlation force is much higher than  the other quantum forces.  The inclusion of such force in the QMHD model, which was neglected in the previous works (e.g., Ref. \onlinecite{sahu2015}), is one of the main purposes of the present study. Furthermore, the source of dissipation is not only the magnetic diffusivity, but also the ion kinematic viscosity which gives an additional term in Eq. \eqref{e32} that was also omitted in the previous studies \cite{sahu2015}.    For typical astrophysical plasmas with $n_0=4\times10^{32}$ m$^{-3}$, $T_e=0.1T_i=5\times10^3$ K and $B_0=5\times10^3$ T, we have $\alpha\sim6\times10^3,~\beta\sim10^{-3},~\varepsilon^2/v_B^2\sim0.3$   and $c_s^2\sim1.5$. Decreasing only the value of $T_e~(\sim10^3$ K) results into a higher value of $\varepsilon^2/v_B^2~(\sim40)$ than 
$c_s^2\sim0.3$. However, slightly decreasing the magnetic field ($B_0=4\times10^3$ T) or increasing the number density ($n_0=7\times10^{32}$ m$^{-3}$)  gives $\alpha\sim9\times10^3,~\beta\sim10^{-3},~\varepsilon^2/v_B^2\sim0.2$   and $c_s^2\sim3$, i.e., higher values of $\alpha$ and $c_s^2$ without any  significant change in  $\beta$ and $\varepsilon^2/v_B^2$.

In  what follows, we numerically solve Eqs.  \eqref{e32} and \eqref{e33}, and  study the influence of the plasma parameters on the large amplitude MSWs. To this end we use MATHEMATICA and apply the finite difference scheme. For a fixed value of the plasma resistivity, the effects of the parameter $\delta$ associated with the kinematic viscosity   on the shock profiles are shown in Fig.
\ref{fig-1}. It is seen that a transition from oscillatory to monotonic shocks occurs with increasing values of $\delta$.  The corresponding phase   portraits are exhibited  in Fig. \ref{figure2}. For very low values of $\delta$, we have a train of oscillations (few of which corresponds to solitons) and the corresponding phase-space trajectory clearly shows a stable closed periodic orbit. As the value of $\delta$ increases, the dissipative effect becomes stronger and the oscillatory shocks tend to  become more and more monotonic.  When the dissipative effect is large enough, we have a completely monotonic shock profile without any oscillation. It is also found that kinematic viscosity has no effect on the amplitude of the shock structures. Thus numerical investigations show the existence of both oscillatory shock for weak dissipation and monotonic shock for strong dissipation.
 Similar features are also observed (not shown in the figure) by increasing the   parameter $\gamma$ associated with the magnetic diffusivity (plasma resistivity) and  keeping $\delta$ fixed. However, in this case, the number of oscillations in front of the shock becomes less in number and the heights of oscillations get reduced. 
\par 
 Figure \ref{fig-3} shows the profiles of MSWs by the effects of the exchange-correlation
parameter $\alpha$. It is observed that as the value of $\alpha$ (or the quantum parameter $H$) decreases, i.e., as one enters into the high-density regimes,  the dissipative effects prevail over that of the quantum particle dispersion,  and the oscillations in front of the shock decrease in number, resulting into the monotonic shock transition. From the   parameter estimation as above, it is also evident that the exchange-correlation force plays a dominating role of dispersion over the other quantum and pressure gradient forces. The individual effects of different quantum forces on the shock profiles can also be stated. If one drops (retains) the term   $\propto$    $\beta$ (however, retains or drops those with $c_s$, $\varepsilon$) and retains the values of $\alpha$, $\gamma$ and $\delta$ as in the upper left panel of Fig. \ref{fig-3}, then only monotonic (oscillatory) shocks can be seen, i.e., the dispersion from the exchange-correlation force is not sufficient to prevail over the dissipation. From the numerical simulation, we also find that  the shock strength decreases with increasing values of $\beta$, however, the same increases (decreases) with increasing (decreasing) values of the  Zeeman energy $\varepsilon$. Thus, we conclude that in a spin QMHD model, one must take into account the  effects of  the exchange-correlation force of fermions along with the quantum force associated with the Bohm potential in order to get more physical insights in the propagation of MSWs in quantum magneto-plasmas. 
\section{Conclusion}
We have presented a theoretical study on the multidimensional propagation of arbitrary amplitude quantum magnetosonic shocks   in a spin-$1/2$ quantum dissipative plasma with the effects of   quantum force (Bohm potential), the pressure-like spin force as well as the exchange and correlation force of electrons. The effects of ion kinematic viscosity and the plasma resistivity are also considered to account for the dissipation in the QMHD model. It is found that the contribution from the exchange-correlation force is dominant over all other similar forces and it plays a significant role on transition from monotonic to oscillatory shocks.  The numerical solution confirms the existence of both oscillatory and
monotonic shock profiles (depending on the strengths of the dissipation and dispersion effects). It is seen that as the ion viscosity or the magnetic diffusivity parameter increases, the oscillatory shock structure becomes more and more monotonic. Also,   both the oscillatory and monotonic shocks depend not only on the dissipative parameters but also on the quantum force (diffraction) or the exchange-correlation force. It is observed that an oscillatory shock profile transforms into a monotonic one when  the value of the quantum diffraction parameter $(\beta)$ (or the particle number density $n_0$) increases or that due to the exchange-correlation force $(\alpha)$ decreases. 
\par
To conclude, the results should be useful for understanding the nonlinear propagation of large amplitude magnetosonic shock-like perturbations that may be generated in many astrophysical   plasma environments such those in the interior of magnetic white dwarf stars, neutron stars etc. where plasma spins up either by means of the plasma viscosity or the interior magnetic field \cite{easson1979}. 

\acknowledgments{APM acknowledges  support from UGC-SAP (DRS, Phase III) with  Sanction  order No.  F.510/3/DRS-III/2015(SAPI),   and UGC-MRP with F. No. 43-539/2014 (SR) and FD Diary No. 3668.}

\end{document}